\documentclass[12pt]{article}
\usepackage[super,compress]{cite}
\usepackage{graphicx}
\usepackage{epsfig}

\newcommand{\be}{\begin{equation}}
\newcommand{\ee}{\end{equation}}

\newcommand{\bea}{\begin{eqnarray}}
\newcommand{\eea}{\end{eqnarray}}
\newcommand{\beq}{\begin{equation}}
\newcommand{\eeq}{\end{equation}}
\newcommand{\nn}{\nonumber}
\def\la{\mathrel{\mathpalette\fun <}}

\def\fun#1#2{\lower3.6pt\vbox{\baselineskip0pt\lineskip.9pt
\ialign{$\mathsurround=0pt#1\hfil##\hfil$\crcr#2\crcr\sim\crcr}}}

\begin{document}

\title{Diquark-diquark-antiquark model for pentaquarks with hidden charm: current status and problems}

\author{V.V. Anisovich$^{+}$, M.A. Matveev$^+$, J. Nyiri$^*$,  A.V. Sarantsev$^+$, A.N. Semenova$^+$,
}
\maketitle

\begin{center}
{\it $^+$Petersburg Nuclear Physics Institute of National Research Centre ''Kurchatov Institute'', Gatchina, 188300, Russia}

{\it $^*$Institute for Particle and Nuclear Physics, Wigner RCP,
Budapest 1121, Hungary}

\end{center}

\begin{abstract}
The $J/\psi p$ signals at 4380 MeV and 4450 MeV which were seen
by the LHCb collaboration in the  decay  $\Lambda^0_b\to K^-J/\psi p$
are discussed following the hypothesis of the existence of diquark-diquark-antiquark composite states.
The discussed problems concern:\\
(i) estimation of masses and mass splittings for low-lying states,\\
(ii) determination of decay modes in the quark recombination scheme,\\
(iii) unitarity and analyticity constrains for pentaquark propagators,  \\
(iv) hadronic deuteron-like components in the diquark-diquark-antiquark multiplet.\\
\end{abstract}



\section{Introduction}
Data for the $\Lambda^0_b\to p J/\psi K^-$ decay provide a definite argument for the existence of a pentaquark, a baryon system in $ p J/\psi$ spectra which has the following quark content: $P_{\bar c cuud}=\bar c(cuud)$.\cite{LHCbB} In terms of the antiquark-diquark-diquark states it can be presented as a three-body system:
\bea \label{2mn}
&&
P=a\;(\bar c^{\alpha}\cdot   D_{cq_1}^{\beta}\cdot D_{q_2q_3}^{\gamma}\epsilon_{\alpha\beta\gamma})
+
b\;(-\bar c^{\alpha}\cdot  D_{q_2q_3}^{\beta}\cdot
D_{cq_1}^{\gamma}\epsilon_{\alpha\beta\gamma}), \nn \\
&&
D_{cq_1}^{\beta} = c_{\beta'}q_{1\beta''}\epsilon^{\beta\beta'\beta''}, \qquad
D_{q_2q_3}^{\gamma} = q_{2\gamma'}q_{3\gamma''}\epsilon^{\gamma\gamma'\gamma''}
\eea
where $\alpha,\beta,\gamma$ refer to color indices. Let us stress that both terms in (\ref{2mn}) have the same weight in color space. The right-hand side diquarks are forming baryons (it is convential condition). The second term in Eq.~(\ref{2mn}) gives a zero result for the states under consideration. Taking into account coordinate wavefunction lead us to nonvanishing second term in Eq.~(\ref{2mn}), however such states are radial excited ones and they are out of our consideration (see Appendix A).

A diquark is a color triplet member, similar to a quark, and the right-hand side of Eq.~(\ref{2mn}) presents a three-body system with a color structure similar to that in low-lying baryons. It was supposed in Refs.~\citen{maia-polo,lebe,ala-2} that it results in the formation of a diquark-diquark-antiquark system. Following this idea it is possible to perform a classification of such baryon states and give estimations of their masses. Estimation of the charmed diquark masses is given in Ref.~\citen{cq-cq} where diquark-antidiquark states with hidden charm are studied.

The notion of the diquark was introduced by Gell-Mann.\cite{gell-mann} Diquarks were discussed for baryon states during a long  time, see the pioneering papers \citen{ida,licht,ono,vva75,schm} and recent studies \citen{santo1,santo2,santo3,santo4,santo5,santo6}. The systematization of baryons in terms of the quark-diquark states is presented in Refs.~\citen{lightdiq,book4}.

Pentaquarks built of light-light and light-heavy diquarks present a natural extension of multiquark schemes studied in the last decade for mesons.\cite{wein,brod,maiani,voloshin,alii,jaff,ross}

The LHCb collaboration observed two resonance states in the $J/\psi \,p$ -spectrum: \cite{LHCbB} the broad state $P(4380)$ and the narrow one $P(4450)$. These resonances generated a wide discussion.\cite{wang,Petrov,espo,os,pol,ali,skw,wo,st,pdg1,pdg2,pdg5,pdg6,pdg7,pdg8,pdg9,pdg10}.

We will concentrate on discussing the narrow state because there are different versions of interpretation for a broad state (for example see Ref.~\citen{odnakartinka}).

We suppose that narrow pentaquark states are formed by decays which are not $s$-wave recombination ones.\cite{ournew} Such a narrow state is $P(4450)$ with the width $\sim 40\pm 20$ MeV. In the present paper we calculate the recombination transitions for multiplet $(\bar ccuud)$. We see that the $P(4450)$ state does not decay in $s$-wave recombination mode. (The $s$-wave transition $P(4450)\to J/\psi + N(J=\frac 32)$ is impossible because of the large mass of $N(J=\frac 32)$ state). Tere is not such a problem for all other multiplet members. The narrow $P(4450)$ recombinates in $d$-vawe $P(4450)\to J/\psi + p$, so the $d$-vawe transition gives a comparatively small value of width.

\section{Pentaquarks as composite states of light diquark ($qq'$), charmed diquark ($qc$) and antiquark ($\bar c$)}

In Ref.~\citen{ala-2} it was supposed that the genuine antiquark-diquark-diquark state in the LHCb experiment is the narrow $P_c^{\frac52-}(4450)$ while the broad $P_c^{\frac32+}(4380)$ is the result of rescatterings of final state hadrons.

\subsection{Color-spin-isospin structure of the pentaquark}

Generally, in color space the wave function of the antiquark-diquark-diquark pentaquark is represented by Eq.~(\ref{2mn}). As it was mentioned above, in the case under consideration we have the first term of Eq.~(\ref{2mn}) only. It is written as follows:
\bea\label{3m}
&&
\bar c^{\alpha}\cdot   D_{cq_1}^{\beta}\cdot D_{q_2q_3}^{\gamma}
\epsilon_{\alpha\beta\gamma}=
{\bar c}^{\alpha}(cq_1)^{\beta}(q_2q_3)^{\gamma}\epsilon_{\alpha\beta\gamma}
\\
&=&-({\bar c}^{\alpha}c_{\alpha})
(q_{1\gamma}q_{2\gamma'}q_{3\gamma''}\epsilon^{\gamma\gamma'\gamma''})+
({\bar c}^{\alpha}
q_{1\alpha})(c_{\gamma}q_{2\gamma'}q_{3\gamma''}\epsilon^{\gamma\gamma'\gamma''})
\nn \\
&=&
-({\bar c}c)(q_{1}q_{2}q_{3})+({\bar c}q_{1})(cq_{2}q_{3})
\nn
\eea
The last line presents the result in a short form, the brackets separate the wite meson and baryon states. To take into account the isospin structure we write:
\be\label{1}
P_{\bar c cuud}= -\sqrt{\frac{1}{3}}\bar c\cdot(cu)\cdot(ud) +\sqrt{\frac{2}{3}}\bar c \cdot(cd)\cdot(uu).
\ee
We discuss a scheme in which the exotic baryon states are formed by standard QCD-motivated interactions (gluonic exchanges, confinement forces) but in addition with diquarks as constituents.

We work with four diquarks, two scalars $S^{II_z,JJ_z}$ and two axial-vectors $A^{II_z,JJ_z}$:
\bea \label{5}
&&
S_{cq}^{\frac 12 I_z,00}(I=1/2,\,J=0),\qquad
A_{cq}^{\frac 12 I_z,1J_z}(I=1/2,\,J=1),
\\
&& S_{q'q''}^{00,00}(I=0,\,J=0),\qquad\,\,
A_{q'q''}^{1I_z,1J_z}(I=1,\,J=1),
\nn
\eea
where $I$ and $J$ refer to isospin and spins of the diquarks. In terms of these diquarks the color-flavor wave funcion of pentaquark reads:
\be
\label{6}
P_{\bar c\cdot cq\cdot q'q''}=\bar c^\alpha\cdot \epsilon_{\alpha\beta\gamma}\,
\begin{tabular}{|l|}
$S^\beta_{cq}$\\
$A^\beta_{cq}$
\end{tabular}
\cdot
\begin{tabular}{|l|}
$S^\gamma_{q'q''}$\\
$A^\gamma_{q'q''}$
\end{tabular}
\ee
We have six diquark-diquark states:
\be  \label{51}
P_{\bar c\cdot cq\cdot q'q''}= \bar c^\alpha\cdot
\begin{tabular}{|l|}
$(S_{cq}S_{q'q''})^\alpha(\frac12;  0^{+})$\\
$(S_{cq}A_{q'q''})^\alpha(\frac12,\frac32;  1^{+})$\\
$(A_{cq}S_{q'q''})^\alpha(\frac12;  1^{+})$\\
$(A_{cq}A_{q'q''})^\alpha(\frac12,\frac32;  0^{+})$\\
$(A_{cq}A_{q'q''})^\alpha(\frac12,\frac32;  1^{+})$\\
$(A_{cq}A_{q'q''})^\alpha(\frac12,\frac32;  2^{+})$
\end{tabular}
\ee
with the isospin and spin-parity numbers of the diquark-diquark pair: $I=\frac12, \frac32$ and $J^P= 0^+, 1^+,2^+$.

So, the low-laing $s$-wave pentaquark multiplet $P(I,J^P)$ reads:
\be\label{7}
P_{\bar c\cdot (cq)\cdot(q'q'')}=
\begin{tabular}{l}
$P(\frac12,\frac12^{-})$\\
$P(\frac12,\frac12^{-})$, $P(\frac12,\frac32^{-})$,
$P(\frac32,\frac12^{-})$, $P(\frac32,\frac32^{-})$, \\
$P(\frac12,\frac12^{-})$, $P(\frac12,\frac32^{-})$\\
$P(\frac12,\frac12^{-})$, $P(\frac32,\frac 12^{-})$\\
$P(\frac12,\frac12^{-})$, $P(\frac12,\frac32^{-})$, $P(\frac32,\frac12^{-})$,
$P(\frac32,\frac32^{-})$
\\
$P(\frac12,\frac32^{-})$, $P(\frac12,\frac52^{-})$,
$P(\frac32,\frac32^{-})$, $P(\frac32,\frac52^{-})$
\end{tabular}
 \ee

\subsection{Masses of the $s$-wave pentaquarks}
It was undersood already relatively long ago that the mass splitting of hadrons can be well described in the framework of the quark model by the short-ranged spin-spin interactions of the constituents.\cite{ZS,RGG,glashow} For mesons and baryons the mass formulae discussed by Glashow read:\cite{glashow}
\bea \label{8a}
&&
M_M=\sum\limits_{j=1,2}m_{q(j)}+a
\frac{\vec{s}_1\vec{s}_2}{m_{q(1)}m_{q(2)}},
\\
&&
M_B=\sum\limits_{j=1,2,3}m_{q(j)}+b
\sum\limits_{j>\ell}\frac{\vec{s}_j\vec{s}_\ell}{m_{q(j)}m_{q(\ell)}},
\nn
\eea
where  $\vec{s}_j$ and  $m_{q(j)}$ refer to spins and masses of the constituents. Mass splitting parameters in Eq.~(\ref{8a}), $a$ and $b$, are characterized by the size of the colour-magnetic interaction and the size of the discussed hadron, the short-range interaction is supposed in Ref.~\citen{GL}. For the 36-plet mesons ($q\bar q$) and 56-plet baryons ($qqq$) formulae of Eq.~(\ref{8a}) work well.

The only exception is the pion, its calculated mass is $\sim$ 350 MeV that point out the existence of additional forces in the pseudoscalar channel (possibly, istanton-induced forces).

It seems natural to apply modified formulae (\ref{8a}) to pentaquark systems. For multiquark states from Ref.~\citen{LHCbB} we have:
\bea \label{8}&&
M_{q_1q_2\cdot q_3c\cdot \bar c}=3m_{D(q_1q_2)D(q_3c)\bar c}
+4\Delta\Big(\vec{\mu}_{D(q_1q_2)}\vec{\mu}_{D(q_3c)}+
\vec{\mu}_{D(q_1q_2)}\vec{\mu}_{\bar c}+\vec{ \mu}_{\bar c}\vec{\mu}_{D(q_3c)}\Big)
\nn \\
&&3m_{D(q_1q_2)D(q_3c)\bar c}=
m_{D(q_1q_2)}+m_{D(q_3c)}+m_{\bar c}
\eea
where $\vec \mu_D$ and $\vec \mu_{\bar c}$ are color-magnetic moments of diquarks and $c$-quark, $\Delta$ is the parameter of spin splitting. Diquarks are considered as composite systems of quarks analogous to light nuclei. The magnetic moments are written as sums of quark magnetic moments:
\bea \label{10e}
&&
\vec{\mu}_{D(q_1q_2)}=\frac{\vec{s}_{q_1} }{m_{q_1}}+
\frac{\vec{s}_{q_2}}{m_{q_2}},
\qquad
\vec{\mu}_{D(q_3c)} = \frac{\vec{s}_{q_3} }{m_{q_3}}
+\frac{\vec{s}_c }{m_{c}}\simeq \frac{\vec{s}_{q_3} }{m_{q_3}},\nn\\
&&
\vec{\mu}_c=\frac{\vec{s}_c }{m_{c}}\simeq 0.
\eea 
In our estimations we take into account that  $m_{c}>>m_{q} $.

Estimation of diquark masses is the most problematic issue in the study of diquarks (see for example Ref.~\citen{santopinto}). Basing on Refs.~\citen{lightdiq,QQQQ} we estimate the masses of scalar $S$ ($J^P=0^+$) and axial $A$ ($J^P=1^+$) diquarks as follows (in MeV units):
\be
\begin{tabular}{ll}
$m_q=330,$ & $m_c=1450$,\\
$m_{S(q_1q_2)}=700$, & $m_{S(q_3c)}=2000$,\\
$m_{A(q_1q_2)}=800$, & $m_{A(q_3c)}=2100$.\\
\end{tabular}
\ee

Correspondingly we write a set of the low-laying pentaquark states:
\be \label{13}
\begin{tabular}{l|l}
           $\qquad I=1/2$&       $\qquad I=3/2 $                          \\
\hline
$P_{\bar c\,A_{(cq)}A_{(q'q'')}}^{(\frac12,\frac32^{-})}(4450)$,
$P_{\bar c\,A_{(cq)}A_{(q'q'')}}^{(\frac12,\frac52^{-})}(4450)$, &
$P_{\bar c\,A_{(cq)}A_{(q'q'')}}^{(\frac32,\frac32^{-})}(4450)$,
$P_{\bar c\,A_{(cq)}A_{(q'q'')}}^{(\frac32,\frac52^{-})}(4450)$\,,\\
$P_{\bar c\,S_{(cq)}S_{(q'q'')}}^{(\frac12,\frac12^{-})}(4300)$, & \\
$P_{\bar c\,S_{(cq)}A_{(q'q'')}}^{(\frac12,\frac12^{-})}(4300)$,
$P_{\bar c\,S_{(cq)}A_{(q'q'')}}^{(\frac12,\frac32^{-})}(4300)$, &
$P_{\bar c\,S_{(cq)}A_{(q'q'')}}^{(\frac32,\frac12^{-})}(4300)$,
$P_{\bar c\,S_{(cq)}A_{(q'q'')}}^{(\frac32,\frac32^{-})}(4300)$, \\
$P_{\bar c\,A_{(cq)}S_{(q'q'')}}^{(\frac12,\frac12^{-})}(4300)$,
$P_{\bar c\,A_{(cq)}S_{(q'q'')}}^{(\frac12,\frac32^{-})}(4300)$, &\\
$P_{\bar c\,A_{(cq)}A_{(q'q'')}}^{(\frac12,\frac12^{-})}(4150)$,
$P_{\bar c\,A_{(cq)}A_{(q'q'')}}^{(\frac12,\frac32^{-})}(4150)$, &
$P_{\bar c\,A_{(cq)}A_{(q'q'')}}^{(\frac32,\frac12^{-})}(4150)$,
$P_{\bar c\,A_{(cq)}A_{(q'q'')}}^{(\frac32,\frac32^{-})}(4150)$, \\
$P_{\bar c\,A_{(cq)}A_{(q'q'')}}^{(\frac12,\frac12^{-})}(4000)$, &
$P_{\bar c\,A_{(cq)}A_{(q'q'')}}^{(\frac32,\frac12^{-})}(4000)$,  \\
\end{tabular}
 \ee
Masses are given in MeV units, an uncertainty in the determination of masses is of the order of $\pm$150 MeV. It should be noted that the process $P\to meson+baryon\to P$ can shift mass values in Eq.~(\ref{13}). Such a rescattering was taken into account for tetraquark states in Ref.~\citen{QQQQ}. Following Ref.~\citen{QQQQ} we estimate this shift to be of the order of 100 MeV.

\section{Spin-isospin structure of pentaquarks }

\subsection {Pentaquark $P_{\bar c A_{cu}\cdot A_{ud} }^{\frac12 \frac12,\frac52 \frac52}(4450)$}

In the antiquark-diquark-diquark scheme the isospin structure of the $P_{\bar ccuud }^{\frac12\frac12,\frac52 \frac52}(4450)$ with spin and isospin projections can be written as follows:
\be\label{n6}
P_{\bar c\uparrow A_{cu}A_{ud}}^{\frac12\frac12,\frac52 \frac52 }(4450)=
-\frac{1}{\sqrt3}\;\bar c^{\uparrow}\cdot A_{cu}^{\frac12\frac12,11}\cdot
A_{ud}^{10,11}
+\sqrt{\frac23}\;\bar c^{\uparrow}\cdot A_{cd}^{\frac12-\frac12,11}\cdot A_{uu}^{11,11}
\ee
The wave functions for axial diquarks read:
\bea\label{n14}
&&A_{cu}^{\frac12\frac12,11}=c^{\uparrow}u^{\uparrow}, \qquad
A_{ud}^{10,11}=\frac{1}{\sqrt2}(d^{\uparrow}u^{\uparrow}+u^{\uparrow}d^{\uparrow}),\nn \\
&&A_{cd}^{\frac12-\frac12,11}=c^{\uparrow}d^{\uparrow}, \qquad
A_{uu}^{11,11}=u^{\uparrow}u^{\uparrow}.
\eea
Taking into account color structure (\ref{3m}), we can rewrite expresion (\ref{n6}) as follows:
\bea
P_{\bar c\uparrow A_{cu}A_{ud}}^{\frac12\frac12,\frac52 \frac52 }(4450)&=&
-\frac{1}{\sqrt3}\;\bar c^{\uparrow}\cdot c^{\uparrow}u^{\uparrow}\cdot
A_{ud}^{10,11}
+\sqrt{\frac23}\;\bar c^{\uparrow}\cdot c^{\uparrow}d^{\uparrow}\cdot A_{uu}^{11,11}=\nn \\
&=&-\frac{1}{\sqrt3}\;\left(-\bar c^{\uparrow}c^{\uparrow}\cdot u^{\uparrow}A_{ud}^{10,11}+\bar c^{\uparrow}u^{\uparrow}\cdot c^{\uparrow}A_{ud}^{10,11}\right)+\nn \\
&+&\sqrt{\frac23}\;\left(-\bar c^{\uparrow}c^{\uparrow}\cdot d^{\uparrow}A_{uu}^{11,11}+\bar c^{\uparrow}d^{\uparrow}\cdot c^{\uparrow}A_{uu}^{11,11}\right).
\eea
The white quark-antiquark and three-quark states should be projected on
states of real mesons and baryons. We need mesons:
\bea\label{n15}
&&
J/\psi^{\Uparrow} =\bar c^{\uparrow}c^{\uparrow},\quad
\bar D^{*0\Uparrow} =\bar c^{\uparrow}u^{\uparrow},\quad
D^{*-\Uparrow} =\bar c^{\uparrow}d^{\uparrow},
\eea
and baryons:
\bea\label{n16}
&&
N^{\frac 12\frac 12,\frac 32\frac 32} =
\sqrt\frac23 d^{\uparrow}A_{uu}^{11,11}-\sqrt\frac13 u^{\uparrow}A_{ud}^{10,11},\nn \\
&&
\Sigma_c^{+(10,\frac32\frac32)}=
c^{\uparrow}A^{10,11}_{ud}\,,\nn \\
&&
\Sigma_c^{++(11,\frac32\frac32)}=
c^{\uparrow}A^{11,11}_{uu}.
\eea
Therefore we write for the pentaquark wave function:
\bea\label{5252}
&&
P_{\bar c\uparrow A_{cu}A_{ud}}^{\frac12\frac12,\frac52 \frac52 }=-
J/\psi^{\Uparrow}N^{\frac 12\frac 12,\frac 32\frac 32}-\sqrt{\frac13}\bar D^{*0\Uparrow}\Sigma^{+(10,\frac32\frac32)}_{c}
+\sqrt{\frac23}D^{*-\Uparrow}\Sigma^{++(11,\frac32\frac32)}_{c}\,.
\eea

The state $P_{\bar c\,A_{(cq)}A_{(q'q'')}}^{(\frac12,\frac52^{-})}(4520\pm 150)$ is a good candidate to be a state which was observed by LHCb \cite{LHCbB}: $5/2^?(4450\pm 4)$ with a width of $\Gamma=39\pm 24$ MeV.

Recombination to $\bar D^{*0}\Sigma^{*+}_{c}$ and $D^{*-\Uparrow}\Sigma^{*++}_{c}$ channels is strongly suppressed because their tresholdls (4460 and 4464 MeV correspondingly) are larger than the pentaquark mass. So the leading recombination channel is $P\to J/\psi + p$ which is possible in $d$-wave only. $s$-wave transition $P\to J/\psi + N(J=\frac 32)$ has the treshold 4817 MeV. That explains the relatively small width of $P(4450)$.

\subsubsection{Propagator}

Resonance propagator:
\bea
&&
\frac{1}{m^2-s +G^2\cos^2\gamma |k_{J/\psi N^*}|
-iG^2\sin^2\gamma k^5_{J/\psi p}-iG^2 k_{D^*\Sigma^*}}
\\
&=&
\frac{1}{m^2_{ P(\frac12 \frac52^-)}-s
-iG^2_{J/\psi p} k^5_{J/\psi p}-iG^2 k_{D^*\Sigma^*}}
\nn
\eea
Here:
\bea
&&
m^2_{ P(\frac12 \frac52^-)}=m^2+G^2\cos^2\gamma |k_{J/\psi N^*}|, \qquad
G^2_{J/\psi p}=G^2\sin^2\gamma
\eea
Thresholds under discussion:
\bea
&&
M_{J/\psi}+M_{N^*}=(3100+1720) MeV=4820\;  MeV,
\\
&&
M_{D^*}+M_{\Sigma^*}=(2010+2520) MeV= 4520\; MeV
\nn\\
&&
M_{J/\psi}+M_{p}=(3100+940) MeV= 4040\; MeV
\nn
\eea

Hadron momenta in the pentaquark region ($\sqrt s\sim 4450 $ MeV):
\bea
&&
|k_{J/\psi N^*}|=\sqrt{\frac{[-s+(M_{J/\psi}+M_{N^*})^2][s-(M_{J/\psi}-M_{N^*})^2]}{4s}}
\\
&&
k_{D^*\Sigma^*}=i\sqrt{\frac{[-s+(M_{D^*}+M_{\Sigma^*})^2][s-(M_{D^*}-M_{\Sigma^*})^2]}{4s}}
,\quad s< (M_{D^*}+M_{\Sigma^*})^2
\nn\\
&&
k_{D^*\Sigma^*}=\sqrt{\frac{[s-(M_{D^*}+M_{\Sigma^*})^2][s-(M_{D^*}-M_{\Sigma^*})^2]}{4s}}
,\quad s> (M_{D^*}+M_{\Sigma^*})^2 \nn\\
\nn
\\
&&
k_{J/\psi p}=\sqrt{\frac{[s-(M_{J/\psi }+M_{ p})^2][s-(M_{J/\psi }-M_{ p})^2]}{4s}}
,\quad s> (M_{J/\psi}+M_{ p})^2
\nn
\eea

Amplitude of production of pentaquark with background contribution
reads:
\bea
A&=&
A_{background}+ \frac{A_{ production}}{m^2_{ P(\frac12 \frac52^-)}-s
-iG^2_{J/\psi p} k^5_{J/\psi p}-iG^2 k_{D^*\Sigma^*}}
\\
&=&
A_{background}\bigg[1+ \frac{\mu^2_{prod}}{m^2_{ P(\frac12 \frac52^-)}-s
-iG^2_{J/\psi p} k^5_{J/\psi p}-iG^2 k_{D^*\Sigma^*}} \bigg]
\nn
\eea
At complex $\mu^2_{prod}$ and $m_{ P(\frac12 \frac52^-)}\sim 4450 \, MeV$
we have resonance-cusp interplay:
\be
\sigma\sim |A|^2
\ee
with
\bea
&&
\mu^2_{prod}=|\mu^2_{prod}|e^{i\Phi_{prod}} \,,\quad
\sqrt{|\mu^2_{prod}|}\la m_{ P(\frac12 \frac52^-)}
\\
&&
\Big[\frac{G^2_{J/\psi p} k^5_{J/\psi p}}{m_{ P(\frac12 \frac52^-)}}
\Big]_{\sqrt s \sim 4500}\quad \la 50\,MeV,\qquad
\Big[
\frac{G^2 k_{D^*\Sigma^*}}{m_{ P(\frac12 \frac52^-)}}
\Big]_{\sqrt s \sim 4500}  \la 50\,MeV
\nn
\eea

\subsection {Pentaquark $ $ $P^{\frac12\frac12,\frac12\frac12}_{\bar c S_{cu}\cdot S_{ud}}$ }
Combination $P=\bar c S_{cu}\cdot S_{ud}$ contains scalar diquarks $S_{cu}^{\frac12\frac12,00}=[c^{\uparrow}u^{\downarrow}-c^{\downarrow}u^{\uparrow}]\frac{1}{\sqrt 2}$ and $S_{ud}^{00,00}=[u^{\uparrow}d^{\downarrow}-u^{\downarrow}d^{\uparrow}-d^{\uparrow}u^{\downarrow}+d^{\downarrow}u^{\uparrow}]\frac{1}{2}$:
\bea\label{n7}
P^{\frac12\frac12,\frac12\frac12}_{\bar c\uparrow(cuud)}&&=\bar c^{\uparrow} S_{cu}^{\frac12\frac12,00}S_{ud}^{10,00}.
\eea
Using (\ref{3m}) expression (\ref{n7}) can be represented as follows:
\bea \label{n1}
P^{\frac12\frac12,\frac12\frac12}_{\bar c\uparrow S_{cu}S_{ud}}&&=
\bar c^{\uparrow}[c^{\uparrow}u^{\downarrow}-c^{\downarrow}u^{\uparrow}]\frac{1}{\sqrt 2}S_{ud}= \\
&&=\frac{1}{\sqrt 2}\left\{-\bar c^{\uparrow}c^{\uparrow}\cdot u^{\downarrow}S_{ud}+\bar c^{\uparrow}u^{\downarrow}\cdot c^{\uparrow}S_{ud}+\bar c^{\uparrow}c^{\downarrow}\cdot u^{\uparrow}S_{ud}-\bar c^{\uparrow}u^{\uparrow}\cdot c^{\downarrow}S_{ud}\right\}.\nn
\eea

Now we need expressions for the meson parts:
\bea\label{n2}
&&\bar c^\uparrow c^{\uparrow}=J/\psi^{\Uparrow},\quad \bar c^\uparrow c^{\downarrow}=\frac{1}{\sqrt2}(J/\psi^{(0)}-\eta_c),\nn \\
&&\bar c^\uparrow u^{\uparrow}=\bar D^{*0\Uparrow},\quad \bar c^\uparrow u^{\downarrow}=\frac{1}{\sqrt2}(\bar D^{*0(0)}-\bar D^{0}),
\eea
and for baryon parts ($\Lambda_c^{+}(cud)$ and $p(uud)$):
\bea\label{n3}
&&\Lambda_c^{+(00,\frac 12\frac 12)}=c^{\uparrow}S^{00,00}_{ud},\\
&&N^{\frac12\frac12,\frac12\frac12}=
 \frac{1}{\sqrt{2}} [\frac{2}{3} d^{\downarrow} A^{11,11}_{uu}
- \frac{\sqrt2}{3} d^{\uparrow} A^{11,10}_{uu}
- \frac{\sqrt2}{3} u^{\downarrow} A^{10,11}_{ud}
+ \frac{1}{3} u^{\uparrow} A^{10,10}_{ud}]
+\frac{1}{\sqrt{2}}u^{\uparrow} S^{00,00}_{ud},\nn\\
&&N^{'\frac12\frac12,\frac12\frac12}=-
 \frac{1}{\sqrt{2}} [\frac{2}{3} d^{\downarrow} A^{11,11}_{uu}
- \frac{\sqrt2}{3} d^{\uparrow} A^{11,10}_{uu}
- \frac{\sqrt2}{3} u^{\downarrow} A^{10,11}_{ud}
+ \frac{1}{3} u^{\uparrow} A^{10,10}_{ud}]
+\frac{1}{\sqrt{2}}u^{\uparrow} S^{00,00}_{ud},\nn
\eea
where $N^{'\frac12\frac12,\frac12\frac12}$ is some radial excitation of proton $N^{\frac12\frac12,\frac12\frac12}$. Taking into account expressions (\ref{n2}) and (\ref{n3}) we can write:
\bea\label{n4}
P^{\frac12\frac12,\frac12\frac12}_{\bar c\uparrow S_{cu}S_{ud}}&&=\frac{1}{\sqrt 2}\left\{J/\psi^{\Uparrow}(p^{\uparrow}+p^{'\uparrow})\frac{1}{\sqrt2}+\frac{1}{\sqrt2}(\bar D^{*0(0)}-\bar D^{0})\Lambda_c^{+\uparrow}+\right.\nn \\
&&\left.+\frac{1}{\sqrt2}(J/\psi^{(0)}-\eta_c)(p^{\uparrow}+p^{'\uparrow})\frac{1}{\sqrt2}-\bar D^{*0\Uparrow}\Lambda_c^{+\downarrow}\right\}.
\eea
Combination $P=-\bar c S_{ud}\cdot S_{cu}$ (see Eq.~(\ref{2mn})) gives zero result without taking into account the coordinate wave function.

\subsection {Pentaquark $P_{\bar c(A_{cu}\cdot A_{ud})_{j=0} }^{\frac12\frac12,\frac12 \frac12 }(\sim 4100 )$}
Isospin and spin structure of pentaquark state $P_{\bar c(A_{cu}\cdot A_{ud})_{j=0} }^{\frac12\frac12,\frac12 \frac12}$ is:
\bea\label{n8}
P_{\bar cA_{cu}\cdot A_{ud}}^{\frac12\frac12,\frac12 \frac12}
&=&
-\frac{1}{\sqrt3}\;\bar c^{\uparrow}\cdot
\frac{1}{\sqrt 3}
[A_{cu}^{\frac12\frac12,11} A_{ud}^{10,1-1}
-A_{cu}^{\frac12\frac12,10} A_{ud}^{10,10}
+A_{cu}^{\frac12\frac12,1-1} A_{ud}^{10,11}]
\nn \\
&&
+\sqrt\frac23\;\bar c^{\uparrow}\cdot
\frac{1}{\sqrt 3}
[A_{cd}^{\frac12-\frac12,11} A_{uu}^{11,1-1}
-A_{cd}^{\frac12-\frac12,10} A_{uu}^{11,10}
+A_{cd}^{\frac12-\frac12,1-1} A_{uu}^{11,11}]
\nonumber
 \\
&=&
-\frac{1}{3}\;\bar c^{\uparrow}\cdot
[c^{\uparrow}u^{\uparrow}A_{ud}^{10,1-1}
-\sqrt\frac12(c^{\uparrow}u^{\downarrow}+c^{\downarrow}u^{\uparrow})A_{ud}^{10,10}
+c^{\downarrow}u^{\downarrow}A_{ud}^{10,11}]
\\
&&
+\frac{\sqrt2}{3}\;\bar c^{\uparrow}\cdot
[c^{\uparrow}d^{\uparrow}A_{uu}^{11,1-1}
-\sqrt\frac12(c^{\uparrow}d^{\downarrow}+c^{\downarrow}d^{\uparrow})A_{uu}^{11,10}
+c^{\downarrow}d^{\downarrow}A_{uu}^{11,11}].
\nonumber
\eea

Using expressions for mesons:
\bea\label{n11}
\bar c^{\uparrow}d^{\uparrow}=D^{*-\Uparrow},\quad \bar c^{\uparrow}d^{\downarrow}=\frac{1}{\sqrt 2}(D^{*-(0)}-D^-),
\eea
and baryons:
\bea\label{n12}
\Sigma_c^{+(10,\frac12\frac12)}=&=&\sqrt{\frac23}c^{\downarrow}A^{10,11}_{ud}-\sqrt{\frac13}c^{\uparrow}A^{10,10}_{ud},\nn \\ 
\Sigma_c^{++(11,\frac12\frac12)}&=&\sqrt{\frac23}c^{\downarrow}A^{11,11}_{uu}-\sqrt{\frac13}c^{\uparrow}A^{11,10}_{uu},
\eea
we can write for the state $P_{\bar c(A_{cu}\cdot A_{ud})_{j=0} }^{\frac12\frac12,\frac12 \frac12}$:
\bea\label{n13}
P_{\bar cA_{cu}\cdot A_{ud}}^{\frac12\frac12,\frac12 \frac12}
&=&\frac12 J/\psi^{\Uparrow}(-N^{\frac12\frac12,\frac12-\frac12}+N^{'\frac12\frac12,\frac12-\frac12})+\nn\\
&+&\frac{1}{2\sqrt 2}(J/\psi^{(0)}-\eta_c)(-N^{\frac12\frac12,\frac12\frac12}+N^{'\frac12\frac12,\frac12\frac12})-\nn\\
&-&\frac{1}{\sqrt 6}\bar D^{*0\Uparrow}\Sigma_c^{+(10,\frac12-\frac12)}-\frac{1}{2\sqrt 3}(\bar D^{*0(0)}-\bar D^0)\Sigma_c^{+(10,\frac12\frac12)}+\nn \\
&+&\frac{1}{\sqrt 3}D^{*-\Uparrow}\Sigma_c^{++(11,\frac12-\frac12)}+\frac{1}{\sqrt 6}(D^{*-(0)}-D^-)\Sigma_c^{++(11,\frac12\frac12)}.
\eea

\subsection {Pentaquark $P_{\bar c A(cu)\cdot A(ud)_{j=1} }^{\frac12\frac12,\frac12\frac12}(\sim 4240 )$}
In the case of diquark-diquark spin $j_{AA}=1$, pentaquark state $P_{\bar c A(cu)\cdot A(ud)}^{\frac12\frac12,\frac12\frac12}$ can be written as follows (upper indecies in right part of the expression represent spin and its projection):
\be
P_{\bar c A(cu)\cdot A(ud)}^{\frac12\frac12,\frac12\frac12}=-\frac{1}{\sqrt3}[\bar c(A_{cu}A_{ud})^{1J_z}]^{\frac12\frac12}+\sqrt{\frac23}[\bar c(A_{cd}A_{uu})^{1J_z}]^{\frac12\frac12}.
\ee
In terms of mesons and baryons it can be represented in form:
\bea
&&P_{\bar c A(cu)\cdot A(ud)}^{\frac12\frac12,\frac12\frac12}=J/\Psi^{\Uparrow}\left(-\frac{1}{3\sqrt2}p^{\downarrow}+\frac{1}{3\sqrt2}p^{'\downarrow}+\frac{1}{3\sqrt2}N^{\frac12\frac12,\frac32-\frac12}\right)+\frac{1}{\sqrt6}J/\Psi^{\Downarrow}N^{\frac12\frac12,\frac32\frac32}+\nn\\
&&+J/\Psi^{(0)}\left(\frac16 p^{\uparrow}-\frac16 p^{'\uparrow}-\frac13 N^{\frac12\frac12,\frac32\frac12}\right)+\eta_c\left(\frac12 p^{\uparrow}-\frac12 p^{'\uparrow}\right)+\nn\\
&&+\bar D^{*0(0)}\left(\frac{1}{3\sqrt6}\Sigma_c^{+10,\frac12\frac12}-\frac{1}{3\sqrt3}\Sigma_c^{+10,\frac32\frac12}\right)+\frac{1}{\sqrt6}\bar D^0\Sigma_c^{+10,\frac12\frac12}+\nn\\
&&+D^{*-(0)}\left(\frac{1}{3\sqrt3}\Sigma_c^{++11,\frac12\frac12}+\frac{2\sqrt2}{3\sqrt3}\Sigma_c^{++11,\frac32\frac12}\right)+D^-\left(-\frac{1}{3\sqrt3}\Sigma_c^{++11,\frac12\frac12}+\frac{\sqrt2}{3\sqrt3}\Sigma_c^{++11,\frac32\frac12}\right)+\nn\\
&&+\bar D^{*0\Uparrow}\left(\frac{1}{3\sqrt6}\Sigma_c^{+10,\frac32-\frac12}-\frac{1}{3\sqrt3}\Sigma_c^{+10,\frac12-\frac12}\right)+\frac{1}{3\sqrt2}\bar D^{*0\Downarrow}\Sigma_c^{+10,\frac32\frac32}-\nn\\
&&-D^{*-\Uparrow}\left(-\frac{\sqrt2}{3\sqrt3}\Sigma_c^{++11,\frac12-\frac12}+\frac{1}{3\sqrt3}\Sigma_c^{++11,\frac32-\frac12}\right)-\frac13 D^{*-\Downarrow}\Sigma_c^{++11,\frac32\frac32}.
\eea

\subsection {Pentaquark $P_{\bar ccuud }^{\frac32 I_z,\frac52 J_z}$ near 4450 MeV}

It is the isotopic counterpart of the observed state
$P_{\bar ccuud }^{\frac12 I_z,\frac52 J_z}(4450)$. We consider here the case with
$I_z=\frac12$, namely, the state with the same quark content as
$P_{\bar ccuud }^{\frac12 \frac12 ,\frac52 J_z}(4450)$. The
$P_{\bar ccuud }^{\frac32\frac12,\frac52 J_z}$
 can be seen in the decay $\Sigma^0_b\to K^-J/\psi\, N\pi$,
in the spectrum $J/\psi\, \Delta^+(1240)\to J/\psi\, N\pi $.

In the quark-diquark model the spin-isospin wave function of the
$P_{\bar ccuud }^{II_z,JJ_z}$
with spin-isospin projections $I_z=\frac12\,,\;J_z=\frac 52$ is written as:
\be
P^{\frac32\frac12, \frac52\frac52}_{\bar ccuud }
=
\sqrt{\frac23}\;\bar c^{\uparrow}\cdot A_{cu}^{\frac12\frac12,11}\cdot
A_{ud}^{10,11}
+\frac{1}{\sqrt3}\;\bar c^{\uparrow}\cdot
A_{cd}^{\frac12-\frac12,11}\cdot A_{uu}^{11,11}
\ee
As previously we use axial-vector diquarks (\ref{n14}) and write:
\bea
P^{\frac32\frac12, \frac52\frac52}_{\bar ccuud }
&=&\sqrt{\frac23}[-\bar c^{\uparrow}c^{\uparrow}\cdot u^{\uparrow}A_{ud}^{10,11}+\bar c^{\uparrow}u^{\uparrow}\cdot c^{\uparrow}A_{ud}^{10,11}]+\nn\\
&+&\frac{1}{\sqrt 3}[-\bar c^{\uparrow}c^{\uparrow}\cdot d^{\uparrow}A_{uu}^{11,11}+\bar c^{\uparrow}d^{\uparrow}\cdot c^{\uparrow}A_{uu}^{11,11}]. 
\eea
The white quark-antiquark and three-quark states should be projected on
states of real mesons (\ref{n15}) and baryons (\ref{n16}), (\ref{26}). Therefore we write for the pentaquark spin-flavor wave function:
\bea \label{25u}
&&
P^{\frac32\frac12, \frac52\frac52}_{\bar ccuud }
=
-J/\psi^{\Uparrow}\Delta^{+(\frac 32\frac 12,\frac 32\frac 32)}
+\sqrt{\frac23}D^{*0\Uparrow}\Sigma^{+(10,\frac 32\frac 32)}_{c}
+\sqrt{\frac13}D^{*-\Uparrow}\Sigma^{++(11,\frac 32\frac 32)}_{c}\,.
\eea
Let us emphasise, that states under consideration are $s$-wave states, we consider them as dominant ones.

\subsection {Pentaquark $P_{\bar cA(cu)A(ud)_{j=1} } ^{\frac12\frac12,\frac32\frac32}$}
Pentaquark state $P_{\bar cA(cu)A(ud)} ^{\frac12\frac12,\frac32\frac32}$ for the case $j_{AA}=1$ can be written as follows:
\bea
&&P_{\bar cA(cu)A(ud)} ^{\frac12\frac12,\frac32\frac32}=-\frac{1}{\sqrt3}\;\Big[\bar c\cdot A_{cu}^{\frac12\frac12,1J_z}\cdot A_{ud}^{10,1J_z}\Big]^{\frac32\frac32}+\sqrt{\frac23}\;\Big[\bar c\cdot A_{cd}^{\frac12-\frac12,1J_z}\cdot A_{uu}^{11,1J_z}\Big]^{\frac32\frac32}=\nn\\
&&=-\frac{1}{\sqrt3}[\bar c^{\uparrow}\frac{1}{\sqrt2}(A_{cu}^{\frac12\frac12,11}A_{ud}^{10,10}-A_{cu}^{\frac12\frac12,10}A_{ud}^{10,11})]+\nn\\
&&+\sqrt{\frac23}[\bar c^{\uparrow}\frac{1}{\sqrt2}(A_{cd}^{\frac12-\frac12,11}A_{uu}^{11,10}-A_{cd}^{\frac12-\frac12,10}A_{uu}^{11,11})].
\eea
Using formulae from \ref{appB} we can rewrite this expression in terms of mesons and baryons:
\bea
&&P_{\bar cA(cu)A(ud)}^{\frac12\frac12,\frac32\frac32}=\frac{1}{\sqrt3}J/\Psi^{\Uparrow}p^{\uparrow}-\frac{1}{\sqrt3}J/\Psi^{\Uparrow}p^{'\uparrow}-\frac{1}{2\sqrt3}J/\Psi^{\Uparrow}N^{\frac12\frac12,\frac32\frac12}+\nn\\
&&+\frac{1}{2\sqrt2}J/\Psi^{(0)}N^{\frac12\frac12,\frac32\frac32}+\frac{1}{2\sqrt2}\eta_cN^{\frac12\frac12,\frac32\frac32}+\frac{\sqrt2}{3}\bar D^{*0\Uparrow}\Sigma_c^{+10,\frac12\frac12}-\frac16\bar D^{*0\Uparrow}\Sigma_c^{+10,\frac32\frac12}-\nn\\
&&-\frac23 D^{*-\Uparrow}\Sigma_c^{+11,\frac12\frac12}+\frac{1}{3\sqrt2}D^{*-\Uparrow}\Sigma_c^{+11,\frac32\frac12}+\frac{1}{2\sqrt6}\bar D^{*0(0)}\Sigma_c^{+10,\frac32\frac32}-\nn\\
&&-\frac{1}{2\sqrt6}\bar D^0\Sigma_c^{+10,\frac32\frac32}-\frac{1}{2\sqrt3}D^{*-(0)}\Sigma_c^{+11,\frac32\frac32}+\frac{1}{2\sqrt3}D^-\Sigma_c^{+11,\frac32\frac32}.
\eea

\subsection {Pentaquark $P_{\bar cA(cu)A(ud) _{j=2}} ^{\frac12\frac12,\frac32\frac32}$}
Pentaquark state $P_{\bar cA(cu)A(ud)} ^{\frac12\frac12,\frac32\frac32}$ for the case $j_{AA}=2$ can be written as follows:
\bea
&&
P_{\bar cA(cu)A(ud)}^{\frac12\frac12,\frac32\frac32}=-\frac{1}{\sqrt3}\;
\Big[\bar c\cdot A_{cu}^{\frac12\frac12,1J_z}\cdot A_{ud}^{10,1J_z}\Big]^{\frac32\frac32}+\sqrt{\frac23}\;\Big[\bar c\cdot A_{cd}^{\frac12-\frac12,1J_z}\cdot A_{uu}^{11,1J_z}\Big]^{\frac32\frac32}.
\eea
In terms of mesons and baryons this state is expressed in the form:
\bea
&&P_{\bar cA(cu)A(ud)}^{\frac12\frac12,\frac32\frac32}=-\frac{\sqrt 5}{2\sqrt 2}\eta_cN^{\frac12\frac12,\frac32\frac32}-\frac{3\sqrt 2}{4\sqrt 5}J/\Psi^{(0)}N^{\frac12\frac12,\frac32\frac32}+\frac{\sqrt 3}{2\sqrt 5}J/\Psi^{\Uparrow}N^{\frac12\frac12,\frac32\frac12}+ \nn\\
&&+\frac{1}{2\sqrt 5}\bar D^{*0\Uparrow}\Sigma_c^{+10,\frac32\frac12}-\frac{\sqrt 6}{4\sqrt 5}\bar D^{*0(0)}\Sigma_c^{+10,\frac32\frac32}-\frac{\sqrt 5}{2\sqrt 6}\bar D^0\Sigma_c^{+10,\frac32\frac32}- \nn\\
&&-\frac{\sqrt 2}{2\sqrt 5}D^{*-\Uparrow}\Sigma_c^{++11,\frac32\frac12}+\frac{\sqrt 5}{2\sqrt 3}D^-\Sigma_c^{++11,\frac32\frac32}+\frac{\sqrt 3}{2\sqrt 5}D^{*-(0)}\Sigma_c^{++11,\frac32\frac32}.
\eea

\subsection {Pentaquark $P_{\bar cS(cu)A(ud) } ^{\frac12\frac12,\frac12\frac12}$}
Using diquarks this state can be represented as follows: 
\bea\label{p1}
&&P_{\bar cS(cu)A(ud)} ^{\frac12\frac12,\frac12\frac12}=
-\frac{1}{\sqrt3}\;
\Big[\bar c\cdot S_{cu}^{\frac12\frac12,00}\cdot A_{ud}^{10,1J_z}\Big]^{\frac12\frac12}
+\sqrt{\frac23}\;\Big[\bar c\cdot S_{cd}^{\frac12-\frac12,00}\cdot A_{uu}^{11,1J_z}\Big]^{\frac12\frac12}=\nn\\
&&=-\frac{1}{\sqrt 3}[\sqrt{\frac23}\bar c^{\downarrow}S_{cu}^{\frac12\frac12,00}A_{ud}^{10,11}-\frac{1}{\sqrt 3}c^{\uparrow}S_{cu}^{\frac12\frac12,00}A_{ud}^{10,10}]+\nn\\
&&+\sqrt{\frac23}[\sqrt{\frac23}\bar c^{\downarrow}S_{cd}^{\frac12-\frac12,00}A_{uu}^{11,11}-\frac{1}{\sqrt 3}c^{\uparrow}S_{cd}^{\frac12-\frac12,00}A_{uu}^{11,10}].
\eea
Using formulae from \ref{appB}, we can rewrite this expression in terms of mesons and baryons:
\bea\label{p2}
&&P_{\bar cS(cu)A(ud)} ^{\frac12\frac12,\frac12\frac12}=-\frac{\sqrt 2}{3}J/\Psi^{(0)}N^{\frac12\frac12,\frac32\frac12}-\frac{\sqrt 2}{12}J/\Psi^{(0)}p^{\uparrow}+\frac{\sqrt 2}{12}J/\Psi^{(0)}p^{'\uparrow}+\frac{\sqrt 2}{4}\eta_c p^{'\uparrow}+\nn\\
&&+\frac{\sqrt 3}{3}J/\Psi^{\Downarrow}N^{\frac12\frac12,\frac32\frac32}+\frac13 J/\Psi^{\Uparrow}N^{\frac12\frac12,\frac32-\frac12}+\frac16 J/\Psi^{\Uparrow}p^{\downarrow}-\frac16 J/\Psi^{\Uparrow}p^{'\downarrow}-\frac{\sqrt 2}{4}\eta_c p^{\uparrow}-\\
&&-\frac{\sqrt 6}{18}\bar D^{*0\Uparrow}\Sigma_c^{+10,\frac12-\frac12}-\frac13\bar D^{*0\Downarrow}\Sigma_c^{+10,\frac32\frac32}+\frac{\sqrt 3}{18}\bar D^{*0(0)}\Sigma_c^{+10,\frac12\frac12}+\frac{\sqrt 3}{6}\bar D^0\Sigma_c^{+10,\frac12\frac12}-\nn\\
&&-\frac{\sqrt 3}{9}\bar D^{*0\Uparrow}\Sigma_c^{+10,\frac32-\frac12}+\frac{\sqrt 6}{9}\bar D^{*0(0)}\Sigma_c^{+10,\frac32\frac12}+\frac{\sqrt 3}{9}D^{*-\Uparrow}\Sigma_c^{++11,\frac12-\frac12}-\frac{\sqrt 6}{6}D^-\Sigma_c^{++11,\frac12\frac12}-\nn\\
&&-\frac{\sqrt 6}{18}D^{*-(0)}\Sigma_c^{++11,\frac12\frac12}+\frac{\sqrt 6}{9}D^{*-\Downarrow}\Sigma_c^{++11,\frac32-\frac12}-\frac{2\sqrt 3}{9}D^{*-(0)}\Sigma_c^{++11,\frac32\frac12}+\frac{\sqrt 2}{3}D^{*-\Downarrow}\Sigma_c^{++11,\frac32\frac32}.\nn
\eea

\subsection {Pentaquark $P_{\bar cA(cu)S(ud) } ^{\frac12\frac12,\frac12\frac12}$}
Here we use axial-vector diquarks $A_{cu}^{\frac12\frac12,11}=c^{\uparrow}u^{\uparrow}$, $A_{cu}^{\frac12\frac12,10}=[c^{\uparrow}u^{\downarrow}+c^{\downarrow}u^{\uparrow}]\frac{1}{\sqrt 2}$ and scalar diquark $S_{ud}^{00,00}=\frac12[u^{\uparrow}d^{\downarrow}-u^{\downarrow}d^{\uparrow}-d^{\uparrow}u^{\downarrow}+d^{\downarrow}u^{\uparrow}]$:
\bea
&&
P_{\bar cA(cu)S(ud) }^{\frac12\frac12,\frac12\frac12}=
\sqrt{\frac23}\bar c^{\downarrow}A_{cu}^{\frac12\frac12,11}S_{ud}^{00,00}-\frac{1}{\sqrt 3}\bar c^{\uparrow}A_{cu}^{\frac12\frac12,10}S_{ud}^{00,00}=\nn\\
&&
\sqrt{\frac23}[-\bar c^{\downarrow}c^{\uparrow}\cdot u^{\uparrow}S_{ud}^{00,00}+\bar c^{\downarrow}u^{\uparrow}\cdot c^{\uparrow}S_{ud}^{00,00}]-\nn\\
&&
\frac{1}{\sqrt 6}[-\bar c^{\uparrow}c^{\uparrow}\cdot u^{\downarrow}S_{ud}^{00,00}+\bar c^{\uparrow}u^{\downarrow}\cdot S_{ud}^{00,00}]-
\frac{1}{\sqrt 6}[-\bar c^{\uparrow}c^{\downarrow}\cdot u^{\uparrow}S_{ud}^{00,00}+\bar c^{\uparrow}u^{\uparrow}\cdot c^{\downarrow}S_{ud}^{00,00}]. 
\eea
Using meson and baryon states we write:
\bea
P_{\bar cA(cu)S(ud)}^{\frac12\frac12,\frac12\frac12}&=&-\frac{1}{\sqrt 6}(J/\psi^{(0)}+\eta_c)(p^{\uparrow}+p^{'\uparrow})+\frac{1}{\sqrt 3}(\bar D^{*0(0)}+D^0)\Lambda_c^{+\uparrow}+\nn\\
&+&\frac{1}{\sqrt 6}J/\psi^{\Uparrow}(p^{\downarrow}+p^{'\downarrow})-\frac{1}{\sqrt 6}(\bar D^{*0(0)}-D^0)\Lambda_c^{+\uparrow}+\nn\\
&+&\frac{1}{\sqrt 6}(J/\psi^{(0)}-\eta_c)(p^{\uparrow}+p^{'\uparrow})-\frac{1}{\sqrt 6}\bar D^{*0\Uparrow}\Lambda_c^{+\downarrow}.
\eea

\section{Conclusion}
We considered the recombination process for pentaquark transition to the meson-baryon channel. With the fixed set of quarks we have two transitions in color space. Both color channels have the same weight. However, different  meson-baryon channels have  different weights due to different phase space and different tresholds. We analized in detail all possible transitions for the system $\bar ccuud$. We suppose that the $s$-wave recombination leads to large decay widths. We see that the $J=\frac 52$ channel only can not decay in the $s$-wave mode; $d$-wave recombination is possible and naturally leads to dumping of width. There is no such dumping of width for other members of pentaquark multiplet with $J=\frac 12,\frac 32$. So we can expect that $P^{\frac 12}$, $P^{\frac 32}$ states are wide and corresponding poles are drowned in complex energy plane. So it is natural that $P(4450)$ is seen experimentally as a narrow peak only.\cite{LHCbB}

\section*{Acknowledgments}
The paper was supported by grant RSF 16-12-10267. We thank A.V. Anisovich for useful discussion.

\appendix

\section{Vanishing of the states $\bar c(q_2q_3)(cq_1)$ }

This state reads:
\bea
-\bar c^{\alpha}\cdot  D_{q_2q_3}^{\beta}\cdot
D_{cq_1}^{\gamma}\epsilon_{\alpha\beta\gamma)}
 &=&
({\bar c}^{\alpha}q_{2\alpha})(q_{3\beta}c_{\beta'}q_{1\beta''}
\epsilon^{\beta\beta'\beta''})-
({\bar c}^{\alpha}q_{3\alpha})(q_{2\beta}c_{\beta'}q_{1\beta''}
\epsilon^{\beta\beta'\beta''})
\nn
\\
&=&({\bar c}q_{2})(q_{3}cq_{1})-({\bar c}q_{3})(q_{2}cq_{1}) =0.
\nn
\eea
The zero results due to flavor-spin symmetry of the light diquark:
$D_{q_2q_3}=D_{q_3q_2}$.
Let us consider as an example combination
$P=\bar c A_{u^{\uparrow}d^{\uparrow}}\cdot A_{c^{\uparrow}u^{\uparrow}} $ :
\bea 
&&
\bar c^{\uparrow}\cdot  A_{u^{\uparrow}d^{\uparrow}}\cdot A_{c^{\uparrow}u^{\uparrow}}
= \bar c^{\uparrow}\cdot \frac{1}{\sqrt 2}( u^{\uparrow}d^{\uparrow}+
d^{\uparrow}u^{\uparrow})\cdot
c^{\uparrow}u^{\uparrow}
\\
&&
= \frac{1}{\sqrt 2} \Big(
\bar c^{\uparrow}\cdot  u^{\uparrow}d^{\uparrow}\cdot  c^{\uparrow}u^{\uparrow}
+
\bar c^{\uparrow}\cdot d^{\uparrow}u^{\uparrow} \cdot  c^{\uparrow}u^{\uparrow}
\Big)
\nn
\\
&&
= \frac{1}{\sqrt 2} \Big(
-\bar c^{\uparrow}u^{\uparrow}\cdot  d^{\uparrow} c^{\uparrow}u^{\uparrow}
+\bar c^{\uparrow}d^{\uparrow}\cdot u^{\uparrow}  c^{\uparrow}u^{\uparrow}
-\bar c^{\uparrow}d^{\uparrow}\cdot  u^{\uparrow} c^{\uparrow}u^{\uparrow}
+\bar c^{\uparrow}u^{\uparrow}\cdot  d^{\uparrow}c^{\uparrow} u^{\uparrow}
\Big)=0.
\nn
\eea

\section{Baryons in terms of diquarks}\label{appB}
\subsection{Proton: $N^{+}(uud)$, $I=\frac12$}
Proton $N^{\frac12\frac12,\frac12J_z}$:
\bea\label{19}
p^{\uparrow}&=&\frac{1}{3\sqrt 2}u^{\uparrow}A^{10,10}_{ud}-\frac{1}{3}u^{\downarrow}A^{10,11}_{ud}-\frac{1}{3}d^{\uparrow}A^{11,10}_{uu}+\frac{\sqrt 2}{3}d^{\downarrow}A^{11,11}_{uu}+\frac{1}{\sqrt 2}u^{\uparrow}S^{00,00}_{ud},\\
p^{\downarrow}&=&-\frac{1}{3\sqrt 2}u^{\downarrow}A^{10,10}_{ud}+\frac{1}{3}u^{\uparrow}A^{10,1-1}_{ud}+\frac{1}{3}d^{\downarrow}A^{11,10}_{uu}-\frac{\sqrt 2}{3}d^{\uparrow}A^{11,1-1}_{uu}-\frac{1}{\sqrt 2}u^{\downarrow}S^{00,00}_{ud}.
\eea

Proton $N^{'\frac12\frac12,\frac12J_z}$:
\bea\label{21}
p^{'\uparrow}&=&-\frac{1}{3\sqrt 2}u^{\uparrow}A^{10,10}_{ud}+\frac{1}{3}u^{\downarrow}A^{10,11}_{ud}+\frac{1}{3}d^{\uparrow}A^{11,10}_{uu}-\frac{\sqrt 2}{3}d^{\downarrow}A^{11,11}_{uu}+\frac{1}{\sqrt 2}u^{\uparrow}S^{00,00}_{ud},\\
p^{'\downarrow}&=&\frac{1}{3\sqrt 2}u^{\downarrow}A^{10,10}_{ud}-\frac{1}{3}u^{\uparrow}A^{10,1-1}_{ud}-\frac{1}{3}d^{\downarrow}A^{11,10}_{uu}+\frac{\sqrt 2}{3}d^{\uparrow}A^{11,1-1}_{uu}-\frac{1}{\sqrt 2}u^{\downarrow}S^{00,00}_{ud}.
\eea

Nucleon $N^{\frac12\frac12,\frac32J_z}$:
\bea\label{23}
N^{\frac12\frac12,\frac32\frac32}&=&-\sqrt{\frac13}u^{\uparrow}A_{ud}^{(10,11)}+\sqrt{\frac23}d^{\uparrow}A_{uu}^{(11,11)},\\
N^{\frac12\frac12,\frac32\frac12}&=&-\frac13 u^{\downarrow}A_{ud}^{(10,11)}-\frac{\sqrt{2}}{3}u^{\uparrow}A_{ud}^{(10,10)}+\frac{\sqrt{2}}{3}d^{\downarrow}A_{uu}^{(11,11)}+\frac23d^{\uparrow}A_{uu}^{(11,10)},\\
N^{\frac12\frac12,\frac32-\frac12}&=&-\frac13 u^{\uparrow}A_{ud}^{(10,1-1)}-\frac{\sqrt{2}}{3}u^{\downarrow}A_{ud}^{(10,10)}+\frac{\sqrt{2}}{3}d^{\uparrow}A_{uu}^{(11,1-1)}+\frac23d^{\downarrow}A_{uu}^{(11,10)},\\
N^{\frac12\frac12,\frac32-\frac32}&=&-\sqrt{\frac13}u^{\downarrow}A_{ud}^{(10,1-1)}+\sqrt{\frac23}d^{\downarrow}A_{uu}^{(11,10)}.
\eea

\subsection{Delta: $\Delta^{+}(uud)$, $I=\frac32$}
Delta $\Delta^{+(\frac32\frac12,\frac12 J_z)}$:
\bea\label{25}
\Delta^{+(\frac32\frac12,\frac12\frac12)}&=&-\frac{1}{3}d^{\uparrow}A^{11,10}_{uu}+\frac{\sqrt2}{3}d^{\downarrow}A^{11,11}_{uu}+\frac{2}{3}u^{\downarrow}A^{10,11}_{ud}-\frac{\sqrt2}{3}u^{\uparrow}A^{10,10}_{ud},\\
\Delta^{+(\frac32\frac12,\frac12-\frac12)}&=&\frac{1}{3}d^{\downarrow}A^{11,10}_{uu}-\frac{\sqrt2}{3}d^{\uparrow}A^{11,1-1}_{uu}-\frac{2}{3}u^{\uparrow}A^{10,1-1}_{ud}+\frac{\sqrt2}{3}u^{\downarrow}A^{10,10}_{ud}.
\eea
Delta $\Delta^{+(\frac32\frac12,\frac32 J_z)}$:
\bea\label{26}
\Delta^{+(\frac32\frac12,\frac32\frac32)}&=&\sqrt{\frac{1}{3}}d^{\uparrow}A^{11,11}_{uu}+\sqrt{\frac{2}{3}}u^{\uparrow}A^{10,11}_{ud},\\
\Delta^{+(\frac32\frac12,\frac32\frac12)}&=&\frac{\sqrt{2}}{3}d^{\uparrow}A^{11,10}_{uu}+\frac{1}{3}d^{\downarrow}A^{11,11}_{uu}+\frac{\sqrt{2}}{3}u^{\downarrow}A^{10,11}_{ud}+\frac{2}{3}u^{\uparrow}A^{10,10}_{ud},\\
\Delta^{+(\frac32\frac12,\frac32-\frac12)}&=&\frac{\sqrt{2}}{3}d^{\downarrow}A^{11,10}_{uu}+\frac{1}{3}d^{\uparrow}A^{11,1-1}_{uu}+\frac{\sqrt{2}}{3}u^{\uparrow}A^{10,1-1}_{ud}+\frac{2}{3}u^{\downarrow}A^{10,10}_{ud},\\
\Delta^{+(\frac32\frac12,\frac32-\frac32)}&=&\sqrt{\frac{1}{3}}d^{\downarrow}A^{11,1-1}_{uu}+\sqrt{\frac{2}{3}}u^{\downarrow}A^{10,1-1}_{ud}.
\eea

\subsection{Lambda: $\Lambda_c^{+}(cud)$, $I=0$}
Lambda $\Lambda_c^{+(00,\frac12 J_z)}$:
\bea\label{27}
\Lambda_c^{+(00,\frac12\frac12)}&=&c^{\uparrow}S^{00,00}_{ud},\\
\Lambda_c^{+(00,\frac12 -\frac12)}&=&c^{\downarrow}S^{00,00}_{ud}.
\eea

\subsection{Sigma: $\Sigma_c^{+}(cud)$, $I=1$}
Sigma $\Sigma_c^{+(10,\frac12J_z)}$:
\bea\label{28}
\Sigma_c^{+(10,\frac12\frac12)}&=&\sqrt{\frac23}c^{\downarrow}A^{10,11}_{ud}-\sqrt{\frac13}c^{\uparrow}A^{10,10}_{ud},\\
\Sigma_c^{+(10,\frac12-\frac12)}&=&-\sqrt{\frac23}c^{\uparrow}A^{10,1-1}_{ud}+\sqrt{\frac13}c^{\downarrow}A^{10,10}_{ud}.
\eea
Sigma $\Sigma_c^{+(10,\frac32J_z)}$:
\bea\label{29}
\Sigma_c^{+(10,\frac32\frac32)}&=&c^{\uparrow}A^{10,11}_{ud},\\
\Sigma_c^{+(10,\frac32\frac12)}&=&\sqrt{\frac13}c^{\downarrow}A^{10,11}_{ud}+\sqrt{\frac23}c^{\uparrow}A^{10,10}_{ud},\\
\Sigma_c^{+(10,\frac32-\frac12)}&=&\sqrt{\frac13}c^{\uparrow}A^{10,1-1}_{ud}+\sqrt{\frac23}c^{\downarrow}A^{10,10}_{ud},\\
\Sigma_c^{+(10,\frac32-\frac32)}&=&c^{\downarrow}A^{10,1-1}_{ud}.
\eea

\subsection{Sigma: $\Sigma_c^{++}(cuu)$, $I=1$}
Sigma $\Sigma_c^{++(11,\frac12J_z)}$:
\bea\label{30}
\Sigma_c^{++(11,\frac12\frac12)}&=&\sqrt{\frac23}c^{\downarrow}A^{11,11}_{uu}-\sqrt{\frac13}c^{\uparrow}A^{11,10}_{uu},\\
\Sigma_c^{++(11,\frac12-\frac12)}&=&-\sqrt{\frac23}c^{\uparrow}A^{11,1-1}_{uu}+\sqrt{\frac13}c^{\downarrow}A^{11,10}_{uu}.
\eea
Sigma $\Sigma_c^{++(11,\frac32J_z)}$:
\bea\label{31}
\Sigma_c^{++(11,\frac32\frac32)}&=&c^{\uparrow}A^{11,11}_{uu},\\
\Sigma_c^{++(11,\frac32\frac12)}&=&\sqrt{\frac13}c^{\downarrow}A^{11,11}_{uu}+\sqrt{\frac23}c^{\uparrow}A^{11,10}_{uu},\\
\Sigma_c^{++(11,\frac32-\frac12)}&=&\sqrt{\frac13}c^{\uparrow}A^{11,1-1}_{uu}+\sqrt{\frac23}c^{\downarrow}A^{11,10}_{uu},\\
\Sigma_c^{++(11,\frac32-\frac32)}&=&c^{\downarrow}A^{11,1-1}_{uu}.
\eea


\begin{thebibliography}{0}
\bibitem{LHCbB}
LHCb Collab. (R.Aaij {\it et al.}), {\it Phys. Rev. Lett.} {\bf 115}, 072001, (2017),\\arXiv:1507.03414v1 [hep-ex].

\bibitem{maia-polo} L.~Maiani, A.D. Polosa and V. Riquer,
{\it Phys.Lett. B} {\bf 749}, 289 (2015), \\arXiv:1507.04980v1 [hep-ph].

\bibitem{lebe}
R. F. Lebed, {\it Phys. Lett. B} {\bf 749}, 454 (2015); arXiv:1507.05867.

\bibitem{ala-2} V.V. Anisovich,  M.A. Matveev, J. Nyiri, A.V. Sarantsev, A.N. Semenova,
{\it Pentaquarks and resonances in the $pJ/\psi$-spectrum },
arXiv:1507.07652v1 [hep-ph], (2015).

\bibitem{cq-cq} V.V. Anisovich,  M.A. Matveev, A.V. Sarantsev, A.N. Semenova,
{\it  Int.J.Mod.Phys. A} {\bf 30} no.32, 1550186 (2015); arXiv:1507.07232 [hep-ph].


\bibitem{gell-mann}M. Gell-Mann,{\it Phys. Lett.} {\bf 8}, 214 (1964).

\bibitem{ida} M.~Ida and R.~Kobayashi,
{\it Progr. Theor. Phys.} {\bf 36}, 846 (1966).

\bibitem{licht}D.B Lichtenberg and L.J.~Tassie, {\it Phys. Rev.} {\bf 155}, 1601 (1967).

\bibitem{ono}S.~Ono, {\it Progr. Theor. Phys.} {\bf 48}, 964 (1972).

\bibitem{vva75} V.V. Anisovich,{\it Pis'ma ZhETF} {\bf 21}, 382 (1975)
[{\it JETP Lett.} {\bf 21}, 174 (1975)];\\
V.V. Anisovich, P.E. Volkovitski, and V.I. Povzun,
{\it ZhETF} {\bf 70}, 1613 (1976) [{\it Sov. Phys. JETP} {\bf 43}, 841 (1976)].

\bibitem{schm}A. Schmidt and R.~Blankenbeckler,
{\it Phys. Rev.} {\bf D16}, 1318 (1977).


\bibitem{santo1}
M. De Sanctis {\it et al.} {\it An interacting quark-diquark model. Strange and nonstrange baryon spectroscopy and other obsevables}, arXiv:1608.00387 [hep-ph].

\bibitem{santo2}J. Ferretti, A. Vassallo and E. Santopinto,{\it Phys. Rev. C} {\bf 83}, 065204 (2011).

\bibitem{santo3} M. De Sanctis {\it et al.},
{\it Eur.Phys.J.} {\bf A52}, 121  (2016); arXiv:1410.0590 [hep-ph].

\bibitem{santo4} E. Santopinto and J. Ferretti, {\it Phys. Rev. C} {\bf 92},  025202 (2015).

\bibitem{santo5} E. Santopinto,{\it Phys. Rev. C} {\bf 72}, 022201 (2005).

\bibitem{santo6} J. Ferretti, A. Vassallo and E. Santopinto, {\it Phys. Rev. C} {\bf 83}, 065204 (2011).

\bibitem{lightdiq}
A.V. Anisovich, V.V. Anisovich, M.A. Matveev, V.A. Nikonov, A.V. Sarantsev, T.O.~Vulfs, {\it  Int. J. Mod. Phys. A} {\bf 25}, 2965 (2010), arXiv:1005.1321 [hep-ph].

\bibitem{book4} A.V. Anisovich, V.V. Anisovich, M.A. Matveev, V.A. Nikonov, J. Nyiri, A.V. Sarantsev {\it Three-particle physics and dispersion relation theory}, (World Scientific, Singapore, 2013).

\bibitem{maiani} L.Maiani, F.Piccinini, A.D.Polosa, V.Riquer, {\it Phys. Rev. D} {\bf 71}, 014028 (2005).

\bibitem{voloshin}M.B. Voloshin, {\it Phys. Rev. D} {\bf 84}, 031502 (2011).

\bibitem{alii}A. Ali, C. Hambrock, W.Wang, {\it Phys. Rev. D} {\bf 85}, 054011 (2012).

\bibitem{wein} S. Weinberg, {\it Phys. Rev. Lett.} {\bf 110}, 261601 (2013).

\bibitem{brod} S.J. Brodsky, D.S. Hwang and R.F. Lebed, {\it Phys. Rev. Lett.} {\bf 113}, 112001 (2014).

\bibitem{jaff} R.L. Jaffe and F. Wilchek, {\it Phys. Rev. Lett.} {\bf 91}, 232003 (2003).

\bibitem{ross} 29 G.C. Rossi and G. Veneziano, {\it Phys. Lett. B} {\bf 597}, 338 (2004).

wang,Petrov,espo,os,pol,ali,skw,wo,st

\bibitem{wang}
Z.-G. Wang, {\it Nucl.Phys. B} {\bf 913}, 163 (2016).

\bibitem{Petrov} M.I. Eides, V.Yu. Petrov, M.V. Polyakov, {\it Phys.Rev. D} {\bf 93}, 054039 (2016).

\bibitem{espo} A. Esposito, A. Pilloni, A.D. Polosa, {\it Phys.Rept.} {\bf 668}, 1 (2016).

\bibitem{os}
E. Oset {\it et al.}, {\it Nucl.Phys. A} {\bf 954}, 371 (2016).

\bibitem{pol}34
I.A. Perevalova, M.V. Polyakov, P. Schweitzer, {\it Phys.Rev. D} {\bf 94}, 054024 (2016).

\bibitem{ali}
A. Ali, I. Ahmed, M.J. Aslam, A. Rehman,  {\it Phys.Rev. D} {\bf 94}, 054001 (2016).

\bibitem{skw}
T. Skwarnicki, Observation of $J/\psi p$ Resonances Consistent with Pentaquark States, in {\it Proc. 27th International Symposium on Lepton Photon Interactions at High Energy (LP15)}, (Ljubljana, Slovenia, August 17-22, 2015), p.~4.

\bibitem{wo}
E. Wang, H.-X. Chen, L.-Sh. Geng, D.-M. Li, E. Oset, {\it Phys.Rev. D} {\bf 93}, 094001 (2016).

\bibitem{st}
Sh. Stone, Exotic hadrons at hadron colliders, {\it Proc. 13th Conference on Flavor Physics and CP Violation (FPCP 2015)}, (Nagoya, Japan, May 25-29, 2015), p.~41.
\bibitem{pdg1}R. Ghosh, A. Bhattacharya, and B. Chakrabarti,{\it The masses
of $P_c$ (4380) and $P_c$ (4450) in the quasi particle diquark model},
arXiv:1508.00356

\bibitem{pdg2} Z.-G. Wang, {\it Eur. Phys. J. C} {\bf 76} (2016), 70, arXiv:1508.01468.

\bibitem{pdg5}J. He, {\it Phys. Lett. B} {\bf 753} (2016) 547, arXiv:1507.05200

 \bibitem{pdg6}Y. Shimizu, D. Suenaga, and M. Harada, {\it Coupled channel analysis of molecule picture of $P_c(4380)$}, arXiv:1603.02376

 \bibitem{pdg7}C.-W. Shen, F.-K. Guo, J.-J. Xie, and B.-S. Zou, {\it Disentangle the hadronic molecule nature of the Pc(4380) pentaquark-like structure},
arXiv:1603.04672

 \bibitem{pdg8}A. Mironov and A. Morozov, {\it JETP Lett.} {\bf 102} (2015), 271, arXiv:1507.04694 [hep-ph].

\bibitem{pdg9}F.-K. Guo, U.-G. Meissner, J. Nieves, Z. Yang,
{\it Remarks on the $P_c$ structures and triangle singularities},
 arXiv:1605.05113 [hep-ph].

\bibitem{pdg10}X.-H. Liu, Q. Wang, Q. Zhao,
 {\it The role of anomalous triangle
singularity in the understanding of the recently observed heavy pentaquark candidates $P_c^+(4380)$ and $P_c^+(4450)$}, AIP Conf.Proc. 1735 (2016) 060004.

\bibitem{odnakartinka}
V.V. Anisovich, M.A. Matveev, A.V. Sarantsev, A.N. Semenova, {\it Modern Physics Letters A} {\bf 30}, 1550212 (2015).

\bibitem{ournew}
V.V. Anisovich, M.A. Matveev, J. Nyiri, A.N. Semenova {\it Modern Physics Letters A} {\bf 32}, 1750154 (2017).


\bibitem{ZS} Y.B. Zeldovich, A.D. Sakharov, Yad. Fis. {\bf 4}, 395 (1966)
[Sov. J. Nucl. Phys. {\bf 4}, 283 (1967)].
\bibitem{RGG} A. de Rujula, H. Georgi, S.L. Glashow, Phys. Rev. D{\bf 12}, 147 (1975).
\bibitem{glashow} S.L. Glashow, {\it Particle physics far from high energy frontier},
Harvard Preprint, HUPT-80/A089 (1980).
\bibitem{GL}
S.L.~Glashow, {\it Particle Physics Far from High Energy Frontier}, (Harvard preprint, HUPT-80/A089, 1980).


\bibitem{santopinto}
E. Santopinto, {\it Phys. Rev. C} {\bf 72}, 022201 (2005).



\bibitem{QQQQ}
V.V. Anisovich, M.A. Matveev, A.V. Sarantsev and A.N. Semenova, {\it Int. J. Mod. Phys. A} {\bf 30}, 1550186 (2015).

\end{thebibliography}
\end{document}